\RequirePackage{lineno} 
\documentclass[aps,prl,twocolumn,showpacs,superscriptaddress]{revtex4}  
\usepackage{graphicx}  
\usepackage{dcolumn}   
\usepackage{bm}        
\usepackage{amssymb}   
\usepackage{amsmath}
\usepackage{multirow}

\begin{document}

\widetext
\vspace{0.2in}


\newcommand{\dzero}     {D\O}
\newcommand{\ppbar}     {\mbox{$p\bar{p}$}}
\newcommand{\ttbar}     {\mbox{$t\bar{t}$}}
\newcommand{\bbbar}     {\mbox{$b\bar{b}$}}
\newcommand{\ccbar}     {\mbox{$c\bar{c}$}}
\newcommand{\pythia}    {{\sc{pythia}}\xspace}
\newcommand{\alpgen}    {{\sc{alpgen}}\xspace}
\newcommand{\geant}     {\sc{geant}}
\newcommand{\met}       {\mbox{$\not\!\!E_T$}}
\newcommand{\rar}       {\rightarrow}
\newcommand{\eps}       {\epsilon}
\newcommand{\bs}        {{\it b $\rightarrow$ s$\gamma$}}
\newcommand{\tb}        {{\it t $\rightarrow$ b}}
\newcommand{\coss}      {\mbox{$\cos\theta^{\star}$}}
\newcommand{\cossp}     {\mbox{\rm{cos}$\theta^{\star}_p$}}
\newcommand{\ptlep}     {$P_T^{lepton}$}
\newcommand{\ljets} {$\ell+$jets}

\hspace{5.2in} \mbox{FERMILAB-PUB-12-043-E}\\

\title{Combination of CDF and D0 measurements of the $\boldsymbol{W}$ boson helicity in top quark decays}
\affiliation{LAFEX, Centro Brasileiro de Pesquisas F\'{i}sicas, Rio de Janeiro, Brazil}
\affiliation{Universidade do Estado do Rio de Janeiro, Rio de Janeiro, Brazil}
\affiliation{Universidade Federal do ABC, Santo Andr\'e, Brazil}
\affiliation{Institute of Particle Physics: McGill University, Montr\'{e}al, Qu\'{e}bec, Canada H3A~2T8; Simon Fraser University, Burnaby, British Columbia, Canada V5A~1S6; University of Toronto, Toronto, Ontario, Canada M5S~1A7; and TRIUMF, Vancouver, British Columbia, Canada V6T~2A3}
\affiliation{University of Science and Technology of China, Hefei, People's Republic of China}
\affiliation{Institute of Physics, Academia Sinica, Taipei, Taiwan 11529, Republic of China}
\affiliation{Universidad de los Andes, Bogot\'a, Colombia}
\affiliation{Charles University, Faculty of Mathematics and Physics, Center for Particle Physics, Prague, Czech Republic}
\affiliation{Czech Technical University in Prague, Prague, Czech Republic}
\affiliation{Center for Particle Physics, Institute of Physics, Academy of Sciences of the Czech Republic, Prague, Czech Republic}
\affiliation{Universidad San Francisco de Quito, Quito, Ecuador}
\affiliation{Division of High Energy Physics, Department of Physics, University of Helsinki and Helsinki Institute of Physics, FIN-00014, Helsinki, Finland}
\affiliation{LPC, Universit\'e Blaise Pascal, CNRS/IN2P3, Clermont, France}
\affiliation{LPSC, Universit\'e Joseph Fourier Grenoble 1, CNRS/IN2P3, Institut National Polytechnique de Grenoble, Grenoble, France}
\affiliation{CPPM, Aix-Marseille Universit\'e, CNRS/IN2P3, Marseille, France}
\affiliation{LAL, Universit\'e Paris-Sud, CNRS/IN2P3, Orsay, France}
\affiliation{LPNHE, Universit\'es Paris VI and VII, CNRS/IN2P3, Paris, France}
\affiliation{CEA, Irfu, SPP, Saclay, France}
\affiliation{IPHC, Universit\'e de Strasbourg, CNRS/IN2P3, Strasbourg, France}
\affiliation{IPNL, Universit\'e Lyon 1, CNRS/IN2P3, Villeurbanne, France and Universit\'e de Lyon, Lyon, France}
\affiliation{III. Physikalisches Institut A, RWTH Aachen University, Aachen, Germany}
\affiliation{Physikalisches Institut, Universit\"at Freiburg, Freiburg, Germany}
\affiliation{II. Physikalisches Institut, Georg-August-Universit\"at G\"ottingen, G\"ottingen, Germany}
\affiliation{Institut f\"ur Experimentelle Kernphysik, Karlsruhe Institute of Technology, D-76131 Karlsruhe, Germany} 
\affiliation{Institut f\"ur Physik, Universit\"at Mainz, Mainz, Germany}
\affiliation{Ludwig-Maximilians-Universit\"at M\"unchen, M\"unchen, Germany}
\affiliation{Fachbereich Physik, Bergische Universit\"at Wuppertal, Wuppertal, Germany}
\affiliation{University of Athens, 157 71 Athens, Greece}
\affiliation{Panjab University, Chandigarh, India}
\affiliation{Delhi University, Delhi, India}
\affiliation{Tata Institute of Fundamental Research, Mumbai, India}
\affiliation{University College Dublin, Dublin, Ireland}
\affiliation{Istituto Nazionale di Fisica Nucleare Bologna, $^{rr}$University of Bologna, I-40127 Bologna, Italy}
\affiliation{Laboratori Nazionali di Frascati, Istituto Nazionale di Fisica Nucleare, I-00044 Frascati, Italy}
\affiliation{Istituto Nazionale di Fisica Nucleare, Sezione di Padova-Trento, $^{ss}$University of Padova, I-35131 Padova, Italy}
\affiliation{Istituto Nazionale di Fisica Nucleare Pisa, $^{oo}$University of Pisa, $^{pp}$University of Siena and $^{qq}$Scuola Normale Superiore, I-56127 Pisa, Italy}
\affiliation{Istituto Nazionale di Fisica Nucleare, Sezione di Roma 1, $^{tt}$Sapienza Universit\`{a} di Roma, I-00185 Roma, Italy}
\affiliation{Istituto Nazionale di Fisica Nucleare Trieste/, I-34100 Trieste, $^{uu}$University of Udine, I-33100 Udine, Italy}
\affiliation{Okayama University, Okayama 700-8530, Japan}
\affiliation{Osaka City University, Osaka 588, Japan}
\affiliation{Waseda University, Tokyo 169, Japan}
\affiliation{University of Tsukuba, Tsukuba, Ibaraki 305, Japan}
\affiliation{Center for High Energy Physics: Kyungpook National University, Daegu 702-701, Korea; Seoul National University, Seoul 151-742, Korea; Sungkyunkwan University, Suwon 440-746, Korea; Korea Institute of Science and Technology Information, Daejeon 305-806, Korea; Chonnam National University, Gwangju 500-757, Korea; Chonbuk National University, Jeonju 561-756, Korea}
\affiliation{Korea Detector Laboratory, Korea University, Seoul, Korea}
\affiliation{CINVESTAV, Mexico City, Mexico}
\affiliation{Nikhef, Science Park, Amsterdam, the Netherlands}
\affiliation{Radboud University Nijmegen, Nijmegen, the Netherlands and Nikhef, Science Park, Amsterdam, the Netherlands}
\affiliation{Joint Institute for Nuclear Research, Dubna, Russia}
\affiliation{Institute for Theoretical and Experimental Physics, Moscow, Russia}
\affiliation{Moscow State University, Moscow, Russia}
\affiliation{Institute for High Energy Physics, Protvino, Russia}
\affiliation{Petersburg Nuclear Physics Institute, St. Petersburg, Russia}
\affiliation{Comenius University, 842 48 Bratislava, Slovakia; Institute of Experimental Physics, 040 01 Kosice, Slovakia}
\affiliation{Institut de Fisica d'Altes Energies, ICREA, Universitat Autonoma de Barcelona, E-08193, Bellaterra (Barcelona), Spain}
\affiliation{Instituci\'{o} Catalana de Recerca i Estudis Avan\c{c}ats (ICREA) and Institut de F\'{i}sica d'Altes Energies (IFAE), Barcelona, Spain}
\affiliation{Centro de Investigaciones Energeticas Medioambientales y Tecnologicas, E-28040 Madrid, Spain}
\affiliation{Instituto de Fisica de Cantabria, CSIC-University of Cantabria, 39005 Santander, Spain}
\affiliation{Stockholm University, Stockholm and Uppsala University, Uppsala, Sweden}
\affiliation{University of Geneva, CH-1211 Geneva 4, Switzerland}
\affiliation{Glasgow University, Glasgow G12 8QQ, United Kingdom}
\affiliation{Lancaster University, Lancaster LA1 4YB, United Kingdom}
\affiliation{University of Liverpool, Liverpool L69 7ZE, United Kingdom}
\affiliation{Imperial College London, London SW7 2AZ, United Kingdom}
\affiliation{University College London, London WC1E 6BT, United Kingdom}
\affiliation{The University of Manchester, Manchester M13 9PL, United Kingdom}
\affiliation{University of Oxford, Oxford OX1 3RH, United Kingdom}
\affiliation{University of Arizona, Tucson, Arizona 85721, USA}
\affiliation{Ernest Orlando Lawrence Berkeley National Laboratory, Berkeley, California 94720, USA}
\affiliation{University of California, Davis, Davis, California 95616, USA}
\affiliation{University of California, Los Angeles, Los Angeles, California 90024, USA}
\affiliation{University of California Riverside, Riverside, California 92521, USA}
\affiliation{Yale University, New Haven, Connecticut 06520, USA}
\affiliation{University of Florida, Gainesville, Florida 32611, USA}
\affiliation{Florida State University, Tallahassee, Florida 32306, USA}
\affiliation{Argonne National Laboratory, Argonne, Illinois 60439, USA}
\affiliation{Fermi National Accelerator Laboratory, Batavia, Illinois 60510, USA}
\affiliation{Enrico Fermi Institute, University of Chicago, Chicago, Illinois 60637, USA}
\affiliation{University of Illinois at Chicago, Chicago, Illinois 60607, USA}
\affiliation{Northern Illinois University, DeKalb, Illinois 60115, USA}
\affiliation{Northwestern University, Evanston, Illinois 60208, USA}
\affiliation{University of Illinois, Urbana, Illinois 61801, USA}
\affiliation{Indiana University, Bloomington, Indiana 47405, USA}
\affiliation{Purdue University Calumet, Hammond, Indiana 46323, USA}
\affiliation{University of Notre Dame, Notre Dame, Indiana 46556, USA}
\affiliation{Purdue University, West Lafayette, Indiana 47907, USA}
\affiliation{Iowa State University, Ames, Iowa 50011, USA}
\affiliation{University of Kansas, Lawrence, Kansas 66045, USA}
\affiliation{Kansas State University, Manhattan, Kansas 66506, USA}
\affiliation{Louisiana Tech University, Ruston, Louisiana 71272, USA}
\affiliation{The Johns Hopkins University, Baltimore, Maryland 21218, USA}
\affiliation{Boston University, Boston, Massachusetts 02215, USA}
\affiliation{Northeastern University, Boston, Massachusetts 02115, USA}
\affiliation{Harvard University, Cambridge, Massachusetts 02138, USA}
\affiliation{Massachusetts Institute of Technology, Cambridge, Massachusetts 02139, USA}
\affiliation{Tufts University, Medford, Massachusetts 02155, USA}
\affiliation{University of Michigan, Ann Arbor, Michigan 48109, USA}
\affiliation{Wayne State University, Detroit, Michigan 48201, USA}
\affiliation{Michigan State University, East Lansing, Michigan 48824, USA}
\affiliation{University of Mississippi, University, Mississippi 38677, USA}
\affiliation{University of Nebraska, Lincoln, Nebraska 68588, USA}
\affiliation{University of New Mexico, Albuquerque, New Mexico 87131, USA}
\affiliation{The Rockefeller University, New York, New York 10065, USA}
\affiliation{Rutgers University, Piscataway, New Jersey 08855, USA}
\affiliation{Princeton University, Princeton, New Jersey 08544, USA}
\affiliation{State University of New York, Buffalo, New York 14260, USA}
\affiliation{Columbia University, New York, New York 10027, USA}
\affiliation{University of Rochester, Rochester, New York 14627, USA}
\affiliation{State University of New York, Stony Brook, New York 11794, USA}
\affiliation{Brookhaven National Laboratory, Upton, New York 11973, USA}
\affiliation{Duke University, Durham, North Carolina 27708, USA}
\affiliation{The Ohio State University, Columbus, Ohio 43210, USA}
\affiliation{Langston University, Langston, Oklahoma 73050, USA}
\affiliation{University of Oklahoma, Norman, Oklahoma 73019, USA}
\affiliation{Oklahoma State University, Stillwater, Oklahoma 74078, USA}
\affiliation{Carnegie Mellon University, Pittsburgh, Pennsylvania 15213, USA}
\affiliation{University of Pittsburgh, Pittsburgh, Pennsylvania 15260, USA}
\affiliation{University of Pennsylvania, Philadelphia, Pennsylvania 19104, USA}
\affiliation{Brown University, Providence, Rhode Island 02912, USA}
\affiliation{University of Texas, Arlington, Texas 76019, USA}
\affiliation{Texas A\&M University, College Station, Texas 77843, USA}
\affiliation{Southern Methodist University, Dallas, Texas 75275, USA}
\affiliation{Rice University, Houston, Texas 77005, USA}
\affiliation{Baylor University, Waco, Texas 76798, USA}
\affiliation{University of Virginia, Charlottesville, Virginia 22904, USA}
\affiliation{University of Washington, Seattle, Washington 98195, USA}
\affiliation{University of Wisconsin, Madison, Wisconsin 53706, USA}
\author{T.~Aaltonen} \affiliation{Division of High Energy Physics, Department of Physics, University of Helsinki and Helsinki Institute of Physics, FIN-00014, Helsinki, Finland}
\author{V.M.~Abazov} \affiliation{Joint Institute for Nuclear Research, Dubna, Russia}
\author{B.~Abbott} \affiliation{University of Oklahoma, Norman, Oklahoma 73019, USA}
\author{B.S.~Acharya} \affiliation{Tata Institute of Fundamental Research, Mumbai, India}
\author{M.~Adams} \affiliation{University of Illinois at Chicago, Chicago, Illinois 60607, USA}
\author{T.~Adams} \affiliation{Florida State University, Tallahassee, Florida 32306, USA}
\author{G.D.~Alexeev} \affiliation{Joint Institute for Nuclear Research, Dubna, Russia}
\author{G.~Alkhazov} \affiliation{Petersburg Nuclear Physics Institute, St. Petersburg, Russia}
\author{A.~Alton$^{a}$} \affiliation{University of Michigan, Ann Arbor, Michigan 48109, USA}
\author{B.~\'{A}lvarez~Gonz\'{a}lez$^b$} \affiliation{Instituto de Fisica de Cantabria, CSIC-University of Cantabria, 39005 Santander, Spain}
% ...
\collaboration{CDF and D0 Collaborations\footnote{With visitors from
$^{a}$Augustana College, Sioux Falls, SD, USA,
$^b$Universidad de Oviedo, E-33007 Oviedo, Spain, 
$^{c}$Northwestern University, Evanston, IL 60208, USA,
$^{d}$The University of Liverpool, Liverpool, UK,
$^e$Universidad Iberoamericana, Mexico D.F., Mexico,
$^{f}$ETH, 8092 Zurich, Switzerland,
$^g$CERN,CH-1211 Geneva, Switzerland,
$^h$Queen Mary, University of London, London, E1 4NS, England,
$^i$National Research Nuclear University, Moscow, Russia,
$^j$Yarmouk University, Irbid 211-63, Jordan,
$^k$Muons, Inc., Batavia, IL 60510,
$^l$UPIITA-IPN, Mexico City, Mexico,
$^{m}$DESY, Hamburg, Germany,
$^{n}$SLAC, Menlo Park, CA, USA,
$^{o}$University College London, London, UK,
$^p$Cornell University, Ithaca, NY  14853, 
$^q$Kansas State University, Manhattan, KS 66506,
$^r$Kinki University, Higashi-Osaka City, Japan 577-8502,
$^s$University of California Santa Barbara, Santa Barbara, CA 93106
$^t$University of Notre Dame, Notre Dame, IN 46556,
$^{u}$Korea University, Seoul, 136-713, Korea,
$^v$Texas Tech University, Lubbock, TX  79609, 
$^w$University of Melbourne, Victoria 3010, Australia,
$^{x}$Centro de Investigacion en Computacion - IPN, Mexico City, Mexico,
$^{y}$Institute of Physics, Academy of Sciences of the Czech Republic, Czech Republic,
$^z$Istituto Nazionale di Fisica Nucleare, Sezione di Cagliari, 09042 Monserrato (Cagliari), Italy,
$^{aa}$University College Dublin, Dublin 4, Ireland,
$^{bb}$University of Iowa, Iowa City, IA  52242,
$^{cc}$University of California Santa Cruz, Santa Cruz, CA  95064,
$^{dd}$ECFM, Universidad Autonoma de Sinaloa, Culiac\'an, Mexico,
$^{ee}$Universidad Tecnica Federico Santa Maria, 110v Valparaiso, Chile,
$^{ff}$University of Cyprus, Nicosia CY-1678, Cyprus, 
$^{gg}$Office of Science, U.S. Department of Energy, Washington, DC 20585, USA,
$^{hh}$CNRS-IN2P3, Paris, F-75205 France,
$^{ii}$Nagasaki Institute of Applied Science, Nagasaki, Japan, 
$^{jj}$Universit{\"a}t Bern, Bern, Switzerland,
$^{kk}$University of California Irvine, Irvine, CA  92697, 
$^{ll}$University of Manchester, Manchester M13 9PL, England,
$^{mm}$University of Fukui, Fukui City, Fukui Prefecture, Japan 910-0017,
and
$^{nn}$Universidade Estadual Paulista, S\~ao Paulo, Brazil.
$^{\ddag}$Deceased.
}}
\noaffiliation

\author{The CDF and D0 Collaborations}
\date{February 22, 2012}

\begin{abstract}
We report the combination of recent measurements of the helicity of the  $W$ boson from top quark decay by the CDF and D0 collaborations, based on data samples corresponding to integrated luminosities of 2.7 -- 5.4~fb$^{-1}$ of $p\bar{p}$ collisions collected during Run II of the Fermilab Tevatron Collider.  Combining measurements that  simultaneously determine the fractions of $W$ bosons with longitudinal ($f_0$) and right-handed ($f_+$) helicities, we find  $f_0 = 0.722 \pm 0.081\mbox{ } [ \pm \ 0.062 \hbox{ (stat.)} \pm  0.052  \hbox{ (syst.)}]$ and $f_+ = -0.033 \pm 0.046\mbox{ }[ \pm \ 0.034 \hbox{ (stat.)} \pm  0.031  \hbox{ (syst.)}]$.  
Combining measurements where one of the helicity fractions is fixed to the value expected in the standard model, we find $f_0 = 0.682 \pm 0.057\mbox{ }[\pm \ 0.035 \hbox{ (stat.)} \pm  0.046  \hbox{ (syst.)}]$ and  $f_+ = -0.015 \pm 0.035\mbox{ }[\pm \ 0.018 \hbox{ (stat.)} \pm  0.030  \hbox{ (syst.)}]$.
The results are consistent with standard model expectations.

\end{abstract}

\pacs{14.65.Ha, 14.70.Fm, 12.15.Ji, 12.38.Qk}
\maketitle


\section{\label{sec:intro}Introduction}
The  study of the properties of the top quark  is one of the major topics of the Tevatron proton-antiproton Collider program at Fermilab.  Using data samples two orders of magnitude larger than were available when the top quark was first observed~\cite{topobs}, the CDF and D0 collaborations have investigated many properties of the top quark, including the helicity of the $W$ bosons produced in the decays $t\rightarrow Wb$.   The on-shell $W$ bosons from  top quark decays can have three possible helicity states, and we denote the fractions of $W^+$ bosons produced in these states as $f_0$ (longitudinal), $f_-$ (left-handed), and $f_+$ (right-handed). In the standard model (SM), the top quark decays via the $V-A$ weak  charged-current interaction, which strongly suppresses right-handed $W^+$ bosons or left-handed $W^-$ bosons (hermitian conjugation will be assumed throughout the paper).  The SM expectation for the helicity fractions depends upon the masses of the top quark ($m_t$) and the $W$ boson ($M_W$).  For the world average values $m_t = 173.3 \pm 1.1$ GeV/$c^2$~\cite{topmass} and  $M_W=80.399 \pm 0.023$ GeV$/c^2$~\cite{wmass}, the expected SM values are $f_0 = 0.688 \pm 0.004$, $f_- =0.310 \pm 0.004$ , and $f_+=0.0017 \pm 0.0001$~\cite{fval2}.  A measurement that deviates significantly from these expectations would provide strong evidence of physics beyond the SM, indicating either a departure from the expected $V-A$ structure of the $tWb$ vertex or the presence of a non-SM contribution to the $t\bar{t}$ candidate sample.  We report  the combination of recent measurements of $f_0$ and $f_+$ from  data recorded at the Tevatron $p\bar{p}$ collider by the CDF and D0 collaborations. The measurements are combined accounting for statistical and systematic
correlations using the method of Refs.~\cite{blue,blue2}.

\section{Input Measurements}

The inputs to the combination are the $f_0$ and $f_+$ values extracted from 2.7 fb$^{-1}$ of CDF data in the lepton~$+$~jets ($t\bar{t}\rightarrow W^+W^-b\bar{b}\rightarrow \ell\nu q\bar{q}^{\prime}b\bar{b}$) channel~\cite{cdfLJwhel} and 5.1 fb$^{-1}$ of CDF data in the dilepton ($t\bar{t}\rightarrow W^+W^-b\bar{b}\rightarrow \ell\nu \ell^\prime\nu b\bar{b}$) channel~\cite{cdfDILwhel} (where $\ell$ and $\ell^\prime$ represent an electron or a muon), and from 5.4 fb$^{-1}$ of D0 data for lepton~$+$~jets and dilepton events analyzed jointly~\cite{d0whel}.  All these measurements use data collected during Run II of the Tevatron.
 Assuming $f_- + f_0 + f_+ = 1$, two types of measurements are performed: (i) a model-independent 
approach where $f_0$ and $f_+$ are determined simultaneously, and (ii) a model-dependent approach where $f_0$ ($f_+$) is fixed 
to its SM value, and  $f_+$ ($f_0$) is measured. The model-independent and model-dependent approaches are referred to as ``2D'' and 
``1D,'' respectively.
  We label the input measurements as follows:
  \begin{itemize}
  \item CDF's measurements of $f_0$ and $f_+$ in the lepton~$+$~jets channel are labeled as $f_{0,{\rm CDF}}^{n{\rm D}, \ell+j}$ and $f_{+,{\rm CDF}}^{n{\rm D}, \ell+j}$, respectively.
  \item CDF's measurements of $f_0$ and $f_+$ in the dilepton channel are labeled as $f_{0,{\rm CDF}}^{n{\rm D}, \ell\ell}$ and $f_{+,{\rm CDF}}^{n{\rm D}, \ell\ell}$, respectively.
  \item D0's measurements of  $f_0$ and $f_+$, which use both the lepton~$+$~jets and dilepton channels,  are labeled as $f_{0,{\rm D0}}^{n{\rm D}}$ and $f_{+,{\rm D0}}^{n{\rm D}}$, respectively.
  \end{itemize}
  Here $n=1$ for 1D measurements and $n=2$ for 2D measurements.
  
The $f_{0(+),{\rm CDF}}^{n{\rm D},\ell+j}$ measurements~\cite{cdfLJwhel}  use the ``matrix element'' method described in Ref.~\cite{memethod}, where the distributions of the momenta of measured jets and leptons as well as the missing transverse energy $\met$  are compared to the expectations for leading-order signal and background matrix elements, convoluted with the detector response to jets and leptons.  The $t\bar{t}$ matrix elements are computed as a function of the $W$ boson helicity fractions to determine the values of $f_0$ and $f_+$ that are most consistent with the data.  

The $f_{0(+),{\rm CDF}}^{n{\rm D},\ell\ell}$ and $f_{0(+),{\rm D0}}^{n{\rm D}}$ measurements are based on the distribution of the helicity angle $\theta^\star$ for each top quark decay, where $\theta^\star$ is the angle in the $W$ boson rest frame between the direction opposite to the top quark and the direction of the down-type fermion (charged lepton or down-type quark) from the decay of the $W$ boson.
  The probability distribution in $\coss$ can be written in terms of the  helicity fractions as follows:
\begin{eqnarray}
\nonumber \omega(\coss) &\propto& 2(1-\cos^2\theta^\star)f_0 + (1-\coss)^2 f_-  \\
                           &             &+ (1+\coss)^2 f_+.
\label{eq:expcost}
\end{eqnarray}
The momentum of the neutrino required to determine  $\theta^\star$ is reconstructed in the lepton~$+$~jets channel through a constrained kinematic fit of each event to the $t\bar{t}$ hypothesis, while for the dilepton channel  $\theta^\star$ is obtained through an algebraic solution of the kinematics.  The distributions in $\coss$ are compared to the expectations from background and $t\bar{t}$ Monte Carlo (MC) simulated events, with different admixtures of helicity fractions, to determine $f_0$ and $f_+$. 

\begin{table*}
\caption{\label{tab:input} Summary of the $W$ boson helicity measurements used in the combination of results.  The CDF measurements have been shifted from their published values to reflect a change in the assumed top quark mass from 175 to 172.5 GeV/$c^2$.  The first uncertainty in brackets below is statistical and the second is systematic.}
\begin{tabular}{cr@{ = }l}
\hline
\hline 
CDF lepton~$+$~jets, 2.7 fb$^{-1}$~\cite{cdfLJwhel}      &  \multicolumn{2}{c}{}  \\
$f_{0,{\rm CDF}}^{2{\rm D},\ell+j}$ & $f_0$&$0.903 \pm 0.123\mbox{ } [\pm \ 0.106 \pm 0.063]$ \\
 $f_{+,{\rm CDF}}^{2{\rm D},\ell+j}$  &   $f_+$&$-0.195 \pm 0.090\mbox{ } [\pm \ 0.067 \pm 0.060]$   \\ \hline
$f_{0,{\rm CDF}}^{1{\rm D},\ell+j}$   &   $f_0$&$0.674 \pm 0.081 \mbox{ } [\pm \ 0.069 \pm 0.042]$    \\
$f_{+,{\rm CDF}}^{1{\rm D},\ell+j}$  &    $f_+$&$-0.044 \pm 0.053\mbox{ } [\pm \ 0.019 \pm 0.050]$  \\ \hline
CDF dilepton, 5.1 fb$^{-1}$  ~\cite{cdfDILwhel}   &  \multicolumn{2}{c}{}  \\
$f_{0,{\rm CDF}}^{2{\rm D},\ell\ell}$ & $f_0$&$0.702 \pm 0.186 \mbox{ }[\pm \ 0.175 \pm 0.062]$             \\
$f_{+,{\rm CDF}}^{2{\rm D},\ell\ell}$           &    $f_+$&$-0.085 \pm 0.096\mbox{ } [\pm \ 0.089 \pm 0.035]$  \\ \hline
$f_{0,{\rm CDF}}^{1{\rm D},\ell\ell}$			&  $f_0$&$0.556 \pm 0.106\mbox{ } [\pm \ 0.088 \pm 0.060]$ 	\\
$f_{+,{\rm CDF}}^{1{\rm D},\ell\ell}$		&  $f_+$&$-0.089 \pm 0.052\mbox{ } [\pm \ 0.041 \pm 0.032]$  \\ \hline	
D0, lepton~$+$~jets and &   \multicolumn{2}{c}{}  \\ 
 dilepton, 5.4 fb$^{-1}$~\cite{d0whel} &   \multicolumn{2}{c}{}  \\
$f_{0,{\rm D0}}^{2{\rm D}}$  & $f_0$&$0.669 \pm 0.102\mbox{ } [\pm \ 0.078 \pm 0.065]$        \\ 
 $f_{+,{\rm D0}}^{2{\rm D}}$                            &  $f_+$&$ 0.023 \pm 0.053\mbox{ } [\pm \ 0.041 \pm 0.034]$      \\ \hline 
 $f_{0,{\rm D0}}^{1{\rm D}}$			&  $f_0$&$0.708 \pm 0.065 \mbox{ }[\pm \ 0.044 \pm 0.048]$ \\
 $f_{+,{\rm D0}}^{1{\rm D}}$			&   $f_+$&$0.010 \pm 0.037 \mbox{ }[\pm \ 0.022 \pm 0.030]$  \\ \hline \hline                             
\end{tabular}
\end{table*}

CDF and D0 treat the top quark mass dependence of the measured helicity fractions differently.   CDF assumes a value of $m_t = 175$ GeV/$c^2$ when reporting central values and includes a description of how the values  change as a function of $m_t$.  D0 assumes a value of $m_t = 172.5$ GeV/$c^2$ and assigns a systematic uncertainty to cover the $m_t$ dependence of the result.  This uncertainty corresponds to a 1.4 GeV/$c^2$ uncertainty on $m_t$, accounting for both the difference between D0's assumed $m_t$ and the world average value and the uncertainty on the world average value~\cite{topmass}.  To facilitate the combination of results, the CDF helicity fractions are shifted to  $m_t$ of 172.5 GeV/$c^2$, and an uncertainty is assigned to account for the 1.4 GeV/$c^2$ uncertainty on $m_t$.  CDF and D0 also use slightly different $M_W$ values in their measurements (80.450 GeV/$c^2$ for CDF and 80.419 GeV/$c^2$ for D0), but this difference changes the expected helicity fractions only by $\approx 10^{-4}$.  The input measurements are summarized in Table~\ref{tab:input}.

\section{Categories of Uncertainty}

The uncertainties on the individual measurements are grouped into categories so that the correlations can be treated properly in the combination.  The categories are specified as follows:

\begin{itemize} 

\item {\bf STA} is the statistical uncertainty.  In each 2D input measurement, there is a strong anticorrelation between the values of $f_0$ and $f_+$, and values of the correlation coefficients are determined from the simultaneous fit for  $f_0$ and $f_+$.  

\item {\bf JES} is the  uncertainty on the jet energy scale.  This uncertainty can arise from theoretical uncertainties on the properties of jets, such as the models for gluon radiation and the fragmentation of $b$ quarks (assessed by comparing the default
 model~\cite{bowler} to an alternative version~\cite{bfragtuning}, and from uncertainties in the calorimeter response.  We assume that the theoretical uncertainties  common to CDF and D0 dominate, and therefore take this uncertainty as fully correlated between CDF and D0.  Details of the jet energy calibration in CDF and D0
 can be found in Refs.~\cite{CDFJES} and~\cite{D0JES}, respectively.

\item {\bf SIG} is the uncertainty on the modeling of $t\bar{t}$ production and decay and has several components.  The effect of uncertainties on the parton distribution functions (PDFs) is estimated using the  $2\times20$ uncertainty sets provided for the CTEQ6M~\cite{cteq6m} PDFs. The uncertainty on the modeling of initial- and final-state gluon radiation is assessed by varying the MC parameters for these processes.    Uncertainties from modeling hadron showers are estimated by
 comparing the expectations from {\sc pythia} and {\sc herwig}.  In addition, D0 estimates the potential impact of next-to-leading order (NLO) effects by comparing the leading-order generators ({\sc alpgen}~\cite{alpgen}, {\sc pythia}~\cite{pythia}, and {\sc herwig}~\cite{herwig}) with the NLO generator 
{\sc mc@nlo}~\cite{mcatnlo},  and 
 the  uncertainty from color reconnection~\cite{color_reconn} by comparing {\sc pythia} models 
with color reconnection turned on and off.   These additional terms increase the $t\bar{t}$ modeling uncertainty by 33\% relative to the value that would be determined using only the components considered in the CDF analyses.  Signal modeling uncertainties
impact the CDF and D0 results in the same manner and therefore are taken as fully correlated among
input measurements.

\item {\bf BGD} is the uncertainty on the modeling of the background.  The procedures used to 
estimate this uncertainty differ for the separate analyses. In CDF's dilepton measurement, the
contribution of each background source is varied within its uncertainty and the resulting effect on
the \coss\ distribution is used to gauge the effect on the measured helicity fractions.  In the CDF lepton~$+$~jets analysis, the change in the result when 
the background is assumed to come from only one source (e.g. only $W+b\bar{b}$ production or
only multijet production), rather than from the expected mixture of sources, is taken as the  uncertainty due to the background shape.  The uncertainty on the background yield is evaluated by varying the assumed signal-to-background ratio. In the D0
measurement, the \coss\ distributions in data and in the background model are compared in a background-dominated sideband region. The background model in the signal region is then reweighted to reflect any differences observed in the background-dominated region, and the resulting changes in the measured helicity fractions are taken as their systematic uncertainties.
The correlations among the background model uncertainties in the input measurements are not known, but are presumably large because of the substantial contribution of $W/Z$ $+$ jets events to the background in each measurement.  We therefore treat this uncertainty as fully correlated between CDF and D0, and also between measurements using dilepton and lepton~$+$~jets events.

\item {\bf MTD} are uncertainties that are specific to a given analysis method.  Effects such as the limitations from the statistics of the MC and any offsets observed in self-consistency tests of the analysis are included in this category.  These uncertainties are fully anticorrelated for 2D measurements of $f_0$ and $f_+$ within a given analysis, but not between different analyses.

\item{\bf MTOP} is the uncertainty due to $m_t$ and is fully correlated between all measurements.

\item{\bf DET} are uncertainties due to the response of the CDF and D0 detectors.   The effects considered include uncertainty in jet energy resolution, lepton identification efficiency, and trigger efficiency.  These uncertainties are found to be negligible in the CDF measurements, but are larger in the D0 measurements due to discrepancies observed in muon distributions between data control samples and MC.  While the cause of these discrepancies was subsequently understood and resolved, D0 
assigns a systematic uncertainty to cover the effect rather than reanalyzing the data.

\item{\bf MHI} is the uncertainty due to multiple hadronic ($p\bar{p}$) interactions in a single bunch crossing.   This uncertainty pertains only to the CDF dilepton measurement.  In D0's measurements the distribution in instantaneous luminosity for the simulated events is reweighted to match that in data, thereby accounting for the impact of multiple interactions.  In CDF's lepton~$+$~jets measurement this uncertainty is found to be negligible.

\end{itemize}

The relationships between the uncertainties reported in individual measurements~\cite{cdfLJwhel,cdfDILwhel,d0whel} and the above categories are given in Table~\ref{tab:errorCat}, and the values of the uncertainties from each input measurement are given in Table~\ref{tab:errorVals}.

\begin{table*}
\caption{\label{tab:errorCat} Relationship between the individual systematic uncertainties on the input measurements~\cite{cdfLJwhel,cdfDILwhel,d0whel} and the categories of uncertainty used for the combination.}
\begin{tabular}{llll}
\hline
\hline 
Uncertainty category      &  \multicolumn{3}{c}{Individual measurement uncertainties}             \\
				     &  CDF lepton~$+$~jets & CDF dilepton  & D0 lepton~$+$~jets and dilepton \\ \hline
JES                            &        Jet energy scale                      &   Jet energy scale    & Jet energy scale \\
                                    &                                                            &                                   &  $b$ fragmentation    \\ \hline
SIG                            &         ISR or FSR                     &    Generators   &  $t\bar{t}$ model \\
                                   &         PDF                     &   ISR or FSR    &   PDF \\
                                   &         Parton shower                    &   PDF            &          \\ \hline
BGD                          &          Background                    &  Background shape    &  Background model \\
                                  &                                &     &   Heavy flavor fraction \\   \hline
MTD                         &    Method-related                             & Template statistics   &   Template statistics \\
                                  &                              &     &    Analysis consistency \\   \hline
MTOP                     &     Top quark mass                          &   Top quark mass   &   Top quark mass \\ \hline
DET                        &                               &      &   Jet energy resolution \\
                                &                               &      &   Jet identification \\
                                &                               &      &    Muon identification  \\
                                &                               &      &    Muon trigger \\ \hline 
MHI                        &                               &    Instant. luminosity &      \\ \hline \hline          
\end{tabular}
\end{table*}

\begin{table*}
\caption{\label{tab:errorVals} Values of the uncertainties from each measurement that are used in the combinations.}
\begin{tabular}{lcccccccc}
\hline
\hline 
Measurement     & STA & JES & SIG & BGD & MTD & MTOP & DET & MHI          \\   \hline
$f_{0,{\rm CDF}}^{2{\rm D},\ell+j}$     &       0.106 & 0.004 & 0.038 & 0.042 & 0.024 & 0.011 & 0.000 & 0.000 \\
$f_{0,{\rm D0}}^{2{\rm D}}$         &       0.078 & 0.011 & 0.039 & 0.032 & 0.022 & 0.009 & 0.031 & 0.000 \\
$f_{0,{\rm CDF}}^{2{\rm D},\ell\ell}$     &      0.175  & 0.002 & 0.050 & 0.023 & 0.028 & 0.005 & 0.000 & 0.013 \\
$f_{+,{\rm CDF}}^{2{\rm D},\ell+j}$      &      0.067  & 0.012 & 0.031 & 0.039 & 0.024 & 0.019 & 0.000 & 0.000 \\
$f_{+,{\rm D0}}^{2{\rm D}}$         &      0.041  & 0.009 & 0.024 & 0.013 & 0.012 & 0.012 & 0.007 & 0.000     \\
$f_{+,{\rm CDF}}^{2{\rm D},\ell\ell}$      &      0.089 & 0.020 & 0.022 & 0.010 & 0.014 & 0.005 & 0.000 & 0.002 \\ \hline 
$f_{0,{\rm CDF}}^{1{\rm D},\ell+j}$      &       0.069 & 0.018 & 0.033 & 0.009 & 0.010 & 0.012 & 0.000 & 0.000 \\
$f_{0,{\rm D0}}^{1{\rm D}}$        &       0.044 & 0.016 & 0.036 & 0.013 & 0.021 & 0.012 & 0.018 & 0.000 \\
$f_{0,{\rm CDF}}^{1{\rm D},\ell\ell}$     &      0.088  & 0.033 & 0.044 & 0.012 & 0.012 & 0.013 & 0.000 & 0.016 \\
$f_{+,{\rm CDF}}^{1{\rm D},\ell+j}$      &      0.019  & 0.017 & 0.024 & 0.038 & 0.005 & 0.015 & 0.000 & 0.000 \\
$f_{+,{\rm D0}}^{1{\rm D}}$        &      0.022  & 0.012 & 0.021 & 0.008 & 0.008 & 0.010 & 0.010 & 0.000     \\
$f_{+,{\rm CDF}}^{1{\rm D},\ell\ell}$      &      0.041 & 0.019 & 0.022 & 0.005 & 0.006 & 0.007 & 0.000 & 0.008 \\ \hline \hline
\end{tabular}
\end{table*}

\section{Combination Procedure}

The results are combined to obtain the best linear unbiased estimators of the correlated observables $f_0$ and $f_+$~\cite{blue2}.    The method uses all the measurements and their covariance matrix ${\bf M}$, where ${\bf M}$ is the sum of the covariance matrices for each category of uncertainty (for the 1D measurements, only the sub-matrices corresponding to the helicity fraction that is varied are relevant):

\begin{eqnarray}
{\bf M} &=& {\bf M}_{\rm STA} +  {\bf M}_{\rm JES}+{\bf M}_{\rm SIG} +{\bf M}_{\rm BGD}\\ 
	& &+{\bf M}_{\rm MTD}+{\bf M}_{\rm MTOP}+{\bf M}_{\rm DET}+{\bf M}_{\rm MHI}. \nonumber
\end{eqnarray}
The correlation coefficients assumed when populating the covariance matrices  for each category of uncertainty are given in Tables~\ref{tab:c_sta}--\ref{tab:c_det}.  When there are correlations in systematic uncertainties between measurements of $f_0$ and $f_+$,  the correlation coefficients are taken to be $-1$, reflecting the large negative statistical correlations observed between measurements of $f_0$ and $f_+$ within a given analysis.

\begin{table}
\caption{\label{tab:c_sta} The statistical correlation coefficients among the measurements used in the combination.}
\begin{tabular} {lr@{.}lr@{.}lr@{.}lr@{.}lr@{.}lr@{.}l}
\hline
\hline 
                      & \multicolumn{2}{c}{$f_{0,{\rm CDF}}^{\ell+j}$ } & \multicolumn{2}{c}{$f_{0,{\rm D0}}$ } & 
                      \multicolumn{2}{c}{$f_{0,{\rm CDF}}^{\ell\ell}$ } & \multicolumn{2}{c}{$f_{+,{\rm CDF}}^{\ell+j}$ } & 
                      \multicolumn{2}{c}{$f_{+,{\rm D0}}$ } & \multicolumn{2}{c}{$f_{+,{\rm CDF}}^{\ell\ell}$ } \\   \hline
$f_{0,{\rm CDF}}^{\ell+j}$   & 1&0               & 0&0              & 0&0                & $-0$&6               & 0&0             & 0&0 \\
$f_{0,{\rm D0}}$     & 0&0               & 1&0              & 0&0                & 0&0                & $-0$&8            & 0&0 \\
$f_{0,{\rm CDF}}^{\ell\ell}$  & 0&0               & 0&0              & 1&0                & 0&0                & 0&0             & $-0$&9 \\
$f_{+,{\rm CDF}}^{\ell+j}$  & $-0$&6              & 0&0              & 0&0                & 1&0                & 0&0             & 0&0  \\
$f_{+,{\rm D0}}$     & 0&0               & $-0$&8            & 0&0                & 0&0                & 1&0              & 0&0 \\
$f_{+,{\rm CDF}}^{\ell\ell}$  & 0&0               & 0&0             & $-0$&9               & 0&0                & 0&0              & 1&0 \\ 
\hline
\hline 
\end{tabular}
\end{table}

\begin{table}
\caption{\label{tab:c_jes} The correlation coefficients among the measurements used in the combination for the JES, SIG, BGD, and MTOP systematic uncertainties.}
\begin{tabular} {lr@{.}lr@{.}lr@{.}lr@{.}lr@{.}lr@{.}l}
\hline
\hline 
                      & \multicolumn{2}{c}{$f_{0,{\rm CDF}}^{\ell+j}$ } & \multicolumn{2}{c}{$f_{0,{\rm D0}}$ } & 
                      \multicolumn{2}{c}{$f_{0,{\rm CDF}}^{\ell\ell}$ } & \multicolumn{2}{c}{$f_{+,{\rm CDF}}^{\ell+j}$ } & 
                      \multicolumn{2}{c}{$f_{+,{\rm D0}}$ } & \multicolumn{2}{c}{$f_{+,{\rm CDF}}^{\ell\ell}$ } \\   \hline
$f_{0,{\rm CDF}}^{\ell+j}$   & 1&0               & 1&0              & 1&0                & $-1$&0               & $-1$&0             & $-1$&0 \\
$f_{0,{\rm D0}}$     & 1&0               & 1&0              & 1&0                & $-1$&0                & $-1$&0            & $-1$&0 \\
$f_{0,{\rm CDF}}^{\ell\ell}$  & 1&0               & 1&0              & 1&0                & $-1$&0                & $-1$&0             & $-1$&0 \\
$f_{+,{\rm CDF}}^{\ell+j}$  & $-1$&0              & $-1$&0              & $-1$&0                & 1&0                & 1&0             & 1&0  \\
$f_{+,{\rm D0}}$     & $-1$&0               & $-1$&0            & $-1$&0                & 1&0                & 1&0              & 1&0 \\
$f_{+,{\rm CDF}}^{\ell\ell}$  & $-1$&0               & $-1$&0            & $-1$&0               & 1&0                & 1&0              & 1&0 \\ 
\hline
\hline 
\end{tabular}
\end{table}

\begin{table}
\caption{\label{tab:c_mtd} The correlation coefficients among the measurements used in the combination for the MTD systematic uncertainty.}
\begin{tabular} {lr@{.}lr@{.}lr@{.}lr@{.}lr@{.}lr@{.}l}
\hline
\hline 
                      & \multicolumn{2}{c}{$f_{0,{\rm CDF}}^{\ell+j}$ } & \multicolumn{2}{c}{$f_{0,{\rm D0}}$ } & 
                      \multicolumn{2}{c}{$f_{0,{\rm CDF}}^{\ell\ell}$ } & \multicolumn{2}{c}{$f_{+,{\rm CDF}}^{\ell+j}$ } & 
                      \multicolumn{2}{c}{$f_{+,{\rm D0}}$ } & \multicolumn{2}{c}{$f_{+,{\rm CDF}}^{\ell\ell}$ } \\   \hline
$f_{0,{\rm CDF}}^{\ell+j}$   & 1&0               & 0&0              & 0&0                & $-1$&0               & $0$&0             & $0$&0 \\
$f_{0,{\rm D0}}$     & 0&0               & 1&0              & 0&0                & $0$&0                & $-1$&0            & $0$&0 \\
$f_{0,{\rm CDF}}^{\ell\ell}$  & 0&0               & 0&0              & 1&0                & $0$&0                & $0$&0             & $-1$&0 \\
$f_{+,{\rm CDF}}^{\ell+j}$  & $-1$&0              & $0$&0              & $0$&0                & 1&0                & 0&0             & 0&0  \\
$f_{+,{\rm D0}}$     & $0$&0               & $-1$&0            & $0$&0                & 0&0                & 1&0              & 0&0 \\
$f_{+,{\rm CDF}}^{\ell\ell}$  & $0$&0               & $0$&0            & $-1$&0               & 0&0                & 0&0              & 1&0 \\ 
\hline
\hline 
\end{tabular}
\end{table}

\begin{table}
\caption{\label{tab:c_det} The correlation coefficients among the measurements used in the combination for the DET and MHI systematic uncertainties.}
\begin{tabular} {lr@{.}lr@{.}lr@{.}lr@{.}lr@{.}lr@{.}l}
\hline
\hline 
                      & \multicolumn{2}{c}{$f_{0,{\rm CDF}}^{\ell+j}$ } & \multicolumn{2}{c}{$f_{0,{\rm D0}}$ } & 
                      \multicolumn{2}{c}{$f_{0,{\rm CDF}}^{\ell\ell}$ } & \multicolumn{2}{c}{$f_{+,{\rm CDF}}^{\ell+j}$ } & 
                      \multicolumn{2}{c}{$f_{+,{\rm D0}}$ } & \multicolumn{2}{c}{$f_{+,{\rm CDF}}^{\ell\ell}$ } \\   \hline
$f_{0,{\rm CDF}}^{\ell+j}$   & 1&0               & 0&0              & 1&0                & $-1$&0               & $0$&0             & $-1$&0 \\
$f_{0,{\rm D0}}$     & 0&0               & 1&0              & 0&0                & $0$&0                & $-1$&0            & $0$&0 \\
$f_{0,{\rm CDF}}^{\ell\ell}$  & 1&0               & 0&0              & 1&0                & $-1$&0                & $0$&0             & $-1$&0 \\
$f_{+,{\rm CDF}}^{\ell+j}$  & $-1$&0              & $0$&0              & $-1$&0                & 1&0                & 0&0             & 1&0  \\
$f_{+,{\rm D0}}$     & $0$&0               & $-1$&0            & $0$&0                & 0&0                & 1&0              & 0&0 \\
$f_{+,{\rm CDF}}^{\ell\ell}$  & $-1$&0               & $0$&0            & $-1$&0               & 1&0                & 0&0              & 1&0 \\ 
\hline
\hline 
\end{tabular}
\end{table}

\section{Results}

\begin{figure}[b]
\includegraphics[scale=0.45]{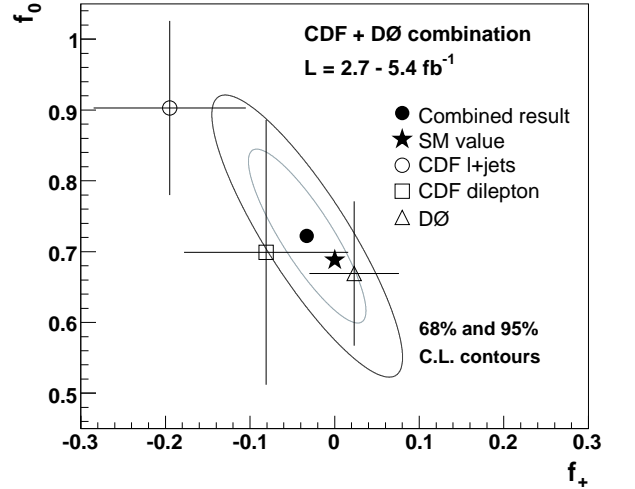}
\caption{\label{fig:cont}  Contours of constant $\chi^2$ for the combination of the 2D  helicity measurements.  The ellipses indicate the 68\% and  95\% C.L.  contours, the dot shows the best-fit value, 
and the star marks the expectation from the SM.  The input measurements to the combination are represented by the open circle, square, and triangle, with error bars indicating the 1$\sigma$ uncertainties on $f_0$ and $f_+$.  Each of the input measurements uses a central value of $m_t = 172.5$ GeV/$c^2$.}
\end{figure}

The result of the combination of the 2D measurements is

\begin{align}
f_0 = 0.722 & \pm  0.081 \\ \nonumber
                    &[ \pm  \ 0.062 \hbox{ (stat.)} \pm  0.052 \hbox{ (syst.)}], \\ \nonumber  
f_+ = -0.033 & \pm   0.046 \\ \nonumber
                     &[ \pm \  0.034 \hbox{ (stat.)} \pm  0.031  \hbox{ (syst.)}].
\end{align}
The contribution from each category of systematic uncertainty is shown in Table~\ref{tab:combsyst}. The combination has a $\chi^2$ value of 8.86 for four degrees of freedom, corresponding to a $p$-value of 6\% for consistency among the input measurements.  The combined values of $f_0$ and $f_+$ have a correlation coefficient of $-0.86$.  The consistency of each input measurement with the combined value and the weight that each input measurement contributes to the combined result  are given in Table~\ref{tab:pulls}. In some cases, the weights have negative values, which can occur in the presence of correlated uncertainties when the most likely value of the observable lies outside of the range of the input measurements~\cite{blue} or, in the case of simultaneous measurements of correlated quantities, when negative weights are needed to satisfy the normalization condition that the weights sum to unity~\cite{blue2}.
Contours of constant   $\chi^2$ in the $f_0$ and $f_+$ plane are shown in Fig.~\ref{fig:cont}.   The SM values for the helicity fractions lie within the 68\% C.L. contour of probability.

\begin{table}
\caption{\label{tab:combsyst} The contribution from each category of systematic uncertainty in the combined measurements. }
\begin{tabular}{lr@{.}lr@{.}lr@{.}lr@{.}l}
\hline
\hline 
Category & \multicolumn{4}{c}{2D combination} &  \multicolumn{4}{c}{1D combination}  \\ 
                   &  \multicolumn{2}{c}{$\delta f_0$}  &  \multicolumn{2}{c}{$\delta f_+$}  &  \multicolumn{2}{c}{$\delta f_0$}  &  \multicolumn{2}{c}{$\delta f_+$} \\ \hline
JES & 0&007   & 0&012 & 0&018 & 0&014 \\
SIG  & 0&038  & 0&022  & 0&036 & 0&021 \\
BGD & 0&028 & 0&013  & 0&012 & 0&009 \\
MTD & 0&014 & 0&008 & 0&007  & 0&006 \\
MTOP & 0&007 & 0&010 & 0&012 & 0&010 \\
DET & 0&016 & 0&003 & 0&011 & 0&007 \\
MHI & 0&001  & 0&0004  & 0&002  & 0&002 \\
\hline
\hline
\end{tabular}
\end{table}

\begin{table}
\caption{\label{tab:pulls} The number of standard deviations of each 2D measurement from the combined values of $f_0$ and $f_+$, and the relative weight contributed by each to the combination.}
\begin{tabular}{lr@{.}lr@{.}lr@{.}l}
\hline
\hline 
Measurement & \multicolumn{2}{c}{s.d. from } &  \multicolumn{2}{c}{Weight for} & \multicolumn{2}{c}{Weight for }   \\ 
                        & \multicolumn{2}{c}{combined values } &   \multicolumn{2}{c}{$f_0$ (\%) } &  \multicolumn{2}{c}{$f_+$ (\%) } \\  \hline
  $f_{0,{\rm CDF}}^{2{\rm D},\ell+j}$         &  1&96   &  44&4                           & $-15$&4  \\
 $f_{0,{\rm D0}}^{2{\rm D}}$           &  $-0$&87    &  45&6                            &  8&1 \\
 $f_{0,{\rm CDF}}^{2{\rm D},\ell\ell}$       & $-0$&12  & 10&1                             &   7&2  \\ \hline
 $f_{+,{\rm CDF}}^{2{\rm D},\ell+j}$        &  $-2$&10   &  27&9                           &   $-3$&6  \\
 $f_{+,{\rm D0}}^{2{\rm D}}$            &  2&06     & $-22$&0                          &   75&9   \\
 $f_{+,{\rm CDF}}^{2{\rm D},\ell\ell}$      & $-0$&62      & $-5$&9                             &  27&7     \\ \hline \hline
\end{tabular}
\end{table}

Combining the 1D measurements yields:

\begin{align}
f_0 = 0.682 & \pm  0.057 \\ \nonumber
                    &[ \pm \ 0.035 \hbox{ (stat.)} \pm  0.046 \hbox{ (syst.)}], \\ \nonumber  
f_+ = -0.015 & \pm   0.035 \\ \nonumber
                     &[ \pm \ 0.018 \hbox{ (stat.)} \pm  0.030  \hbox{ (syst.)}].
\end{align}

The contribution of each category of systematic uncertainty is shown in Table~\ref{tab:combsyst}. The combination for $f_0$ ($f_+$) has a $\chi^2$ of 2.12  (4.44) for two degrees of freedom,
corresponding to a $p$-value of 35\% (11\%) for consistency among the input measurements.
  The consistency of each input measurement with the combined value  and the weight that each input measurement contributes to the combined result are presented in Table~\ref{tab:pulls1D}.
  
 \begin{table}
\caption{\label{tab:pulls1D} The number of standard deviations of each 1D measurement from the combined values of $f_0$ and $f_+$, and the relative weight contributed by each to the combination.}
\begin{tabular}{lr@{.}lr@{.}l}
\hline
\hline 
Measurement & \multicolumn{2}{c}{s.d. from } &  \multicolumn{2}{c}{Weight (\%)} \\
                           & \multicolumn{2}{c}{combined values} \\ \hline
  $f_{0,{\rm CDF}}^{1{\rm D},\ell+j}$      &  $-0$&15    &  31&3            \\
 $f_{0,{\rm D0}}^{1{\rm D}}$           &  0&83     &  58&9         \\
 $f_{0,{\rm CDF}}^{1{\rm D},\ell\ell}$        & $-1$&40     & 9&9    \\ \hline
 $f_{+,{\rm CDF}}^{1{\rm D},\ell+j}$        &  $-0$&70   &  5&2     \\
 $f_{+,{\rm D0}}^{1{\rm D}}$           &  2&08     &  72&1      \\
 $f_{+,{\rm CDF}}^{1{\rm D},\ell\ell}$      & $-1$&94   &  22&7         \\ \hline \hline
\end{tabular}
\end{table}

 
\section{\label{sec:summ}Summary}

We have combined measurements of the  helicity of $W$ bosons arising from top quark decay in \ttbar\ events from the CDF and D0 collaborations, finding
\begin{align}
f_0 = 0.722 & \pm  0.081 \\ \nonumber
                    &[ \pm \  0.062 \hbox{ (stat.)} \pm  0.052 \hbox{ (syst.)}], \\ \nonumber  
f_+ = -0.033 & \pm   0.046 \\ \nonumber
                     &[ \pm \ 0.034 \hbox{ (stat.)} \pm  0.031  \hbox{ (syst.)}]
\end{align}
for measurements in which both $f_0$ and $f_+$ are varied simultaneously, and
\begin{align}
f_0 = 0.682 & \pm  0.057 \\ \nonumber
                    &[ \pm \ 0.035 \hbox{ (stat.)} \pm  0.046 \hbox{ (syst.)}], \\ \nonumber  
f_+ = -0.015 & \pm   0.035 \\ \nonumber
                     &[ \pm \ 0.018 \hbox{ (stat.)} \pm  0.031  \hbox{ (syst.)}]
\end{align}
when one of the helicity fractions is held fixed at the SM value.

These are the most precise measurements of $f_0$ and $f_+$ to date. The results are consistent with expectations from the SM and provide no indication of new physics in the $tWb$ coupling nor of the presence of a non-SM source of events in the selected sample.

\section{Acknowledgement}

We thank the staffs at Fermilab and collaborating institutions,
and acknowledge support from the
DOE and NSF (USA);
CEA and CNRS/IN2P3 (France);
FASI, Rosatom and RFBR (Russia);
CNPq, FAPERJ, FAPESP and FUNDUNESP (Brazil);
DAE and DST (India);
INFN (Italy);
Ministry of Education, Culture,
Sports, Science and Technology (Japan);
Colciencias (Colombia);
CONACyT (Mexico);
World Class University Program, National Research Foundation, NRF (Korea);
CONICET and UBACyT (Argentina);
Australian Research Council (Australia);
FOM (The Netherlands);
STFC and the Royal Society (United Kingdom);
MSMT and GACR (Czech Republic);
CRC Program and NSERC (Canada);
Academy of Finland (Finland);
BMBF and DFG (Germany);
SFI (Ireland);
Slovak R\&D Agency (Slovakia);
Programa Consolider-Ingenio 2010 (Spain);
The Swedish Research Council (Sweden);
Swiss National Science Foundation (Switzerland);
NSC (Republic of China);
CAS and CNSF (China) and 
the A.P. Sloan Foundation (USA).

\end{document}